\documentclass[pre,english,showpacs,twocolumn,aps,a4paper,floatfix]{revtex4}
\usepackage[T1]{fontenc}
\usepackage{color}
\usepackage[latin1]{inputenc}
\usepackage{graphicx}
\usepackage{bm}
\usepackage{epsfig}
\usepackage{stmaryrd}

\begin{document}

\title{Hard rods in a cylindrical pore: the nematic-to-smectic phase transition}

\author{Szabolcs Varga}

\email{vargasz@almos.uni-pannon.hu}

\affiliation{Institute of Physics and Mechatronics, University of Pannonia, PO Box 158, Veszpr\'em, H-8201 Hungary}

\author{Yuri Mart\'{\i}nez-Rat\'on}

\email{yuri@math.uc3m.es}

\affiliation{Grupo Interdisciplinar de Sistemas Complejos (GISC), Departamento de Matem\'aticas, Escuela Polit\'ecnica Superior,
Universidad Carlos III de Madrid, Avenida de la Universidad 30, E-28911, Legan\'es, Madrid, Spain}

\author{Enrique Velasco}

\email{enrique.velasco@uam.es}

\affiliation{Departamento de F\'{\i}sica Te\'orica de la Materia Condensada, Instituto de Ciencia de
Materiales Nicol\'as Cabrera and Condensed Matter Physics Center (IFIMAC), Universidad Aut\'onoma de Madrid, E-28049 Madrid, Spain}

\begin{abstract}
The effect of cylindrical confinement on the phase behaviour of a system of parallel hard rods is studied using Onsager's 
second virial theory. The hard rods are represented as hard cylinders of diameter $D$ and length $L$, while the cylindrical
pore is infinite with diameter $W$. The interaction between the wall and the rods is hard repulsive, and it is assumed that 
molecules are parallel to the surface of the pore (planar anchoring). In very narrow pores ($D<W<2 D$), the structure is 
homogeneous and the system behaves as a one-dimensional Tonks gas. For wider pores, inhomogeneous fluid structures emerge 
because of the lowering of the average excluded volume due to the wall-particle interaction. The bulk nematic-smectic A phase
transition is replaced by a transition between inhomogeneous nematic and smectic A phases. The smectic 
is destabilized with respect to the nematic for decreasing pore width; this effect becomes substantial for $W<10 D$. 
For $W>100 D$, results for bulk and confined fluids agree well due to the short range effect of the wall ($\sim 3-4D$).
\end{abstract}

\pacs{61.30.Pq,64.70.M-,47.57.J-}

\maketitle

\section{Introduction}

Confined liquid crystals have received considerable experimental and theoretical attention over the past decades due to their 
fundamental and technological importance. Liquid crystals are in contact with surfaces in many devices such as liquid-crystal 
(LC) displays, optical shutters and LC thermometers. In these systems it is very important to know the effect of geometrical 
confinement, surface roughness and particle-surface interactions on the ordering properties of LC materials \cite{1}.
Flat and curved surfaces, confinement between parallel surfaces, tubular pores, nanochannels and porous materials are just a 
few examples where the phase behaviour of the LCs is modified \cite{2,3,4,5,6,7,8,9,10}. Several phenomena have been 
observed in confined LCs such as surface phase transitions \cite{11}, capillary nematization \cite{12}, layering transitions \cite{13} 
and suppression of phase transitions \cite{14,15}. In addition to calamitic LCs, the ordering properties of discotic 
LCs have also been examined in the presence of different confinements \cite{16,17}. Even binary mixtures of plates
and rods have been studied at a single wall and in a slit pore \cite{18}.

The objective of our work is to examine the effect of the cylindrical confinement on the phase behaviour and structural 
properties of parallel hard cylinders, assuming that the cylinders are parallel to the surface of the pore (planar anchoring) with 
their axes pointing along the cylinder axis. It has been found that a bulk system of parallel hard cylinders has three 
different phases, namely, nematic, smectic A and solid \cite{19,20}. Here we focus only on the nematic and smectic A (hereafter
called simply `smectic') phases, while 
the solid phase is left for future studies. We use the well-known Onsager second virial theory \cite{21,22} which gives reasonable 
results for the nematic and smectic phases of parallel hard rods \cite{23}. In our previous work we have shown that Onsager theory
 can be applied successfully to study the nematic and smectic ordering of homeotropically anchored uniaxial and biaxial hard rods 
in slit-like pores \cite{24,25}. In particular, we showed \cite{24} that the nematic-smectic transition is suppressed due to the 
restricted geometry along the direction of the density modulation. In the present work we change our focus to the cylindrical 
confinement and consider planar anchoring. This system behaves very differently from the slit-like pore: Since the cylinder is 
infinite in the direction of the smectic density wave, there are no geometric restrictions and the nematic-smectic phase 
transition can survive in the cylindrical pore. Our study reveals that planar anchoring widens the stability window of the 
nematic phase and shrinks the region of smectic ordering in cylindrical pores. The case of hometropic anchoring in cylindrical 
geometry involves topological restrictions that give rise to defects and is not considered here.

\section{Theory}
\begin{figure}
\epsfig{file=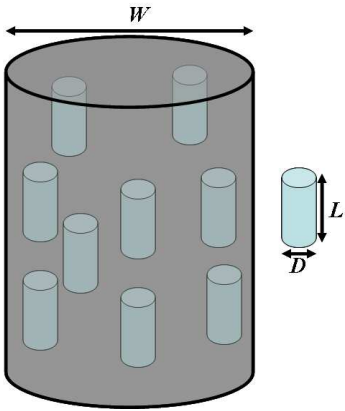,width=2.in}
\caption{
Schematic of the confinement and the molecular model:
$W$ is the width of the cylindrical pore, and $L$ and $D$ are the length and the diameter of the hard cylinder particle, 
respectively.}
\label{fig1}
\end{figure}

We study the phase behaviour of a system of parallel hard cylinders confined in a cylindrical hard pore, 
as shown in Fig. \ref{fig1} 
The molecular parameters are the length ($L$) and the diameter ($D$) of the cylinder. The particle orientations are taken to be
 parallel to the cylindrical pore, which is infinitely long but has a finite size in the perpendicular plane. Note that the 
width of the pore ($W$) must be larger than $D$. The appropriate thermodynamic function for studying confined fluids is the 
grand-potential density functional $\Omega[\rho]$ [26], which is given by
\begin{eqnarray}
 \beta\Omega[\rho]=\beta {\cal F}[\rho]-\int_V d{\bm r}\rho({\bm r})\left[\beta\mu-\beta V_{\hbox{\tiny ext}}({\bm r})\right],
\label{f1}
\end{eqnarray}
where $\beta=1/kT$, $k$ is Boltzmann's constant, $T$ is temperature, 
${\cal F}$ is the Helmholtz free-energy functional, $V$ is the volume of the 
pore, ${\bm r}$ is the position vector of the rod, $\rho({\bm r})$ is the local number density, and $\mu$ is the chemical
potential. We attach a cylindrical coordinate frame to the centre of the pore.
Let $(R,\phi,z)$ be the cylindrical coordinates of a position vector ${\bm r}$. The external potential due to the confinement, 
$V_{\hbox{\tiny ext}}({\bm r})$, is considered to be hard, i.e. the surfaces of the cylinders have to be inside the pore. 
It takes the form
\begin{eqnarray}
 \beta V_{\hbox{\tiny ext}}({\bm r})=\left\{\begin{array}{ll}\infty,&R>R_e\\\\0,&0\le R\le R_e,\end{array}\right.
\label{f2}
\end{eqnarray}
where $R_e=(W-D)/2$ is the effective pore radius. The Helmholtz free energy, which is the sum of ideal and excess contributions 
(${\cal F}={\cal F}_{\hbox{\tiny id}}+{\cal F}_{\hbox{\tiny ex}}$), can be determined from
\begin{eqnarray}
&& \beta{\cal F}_{\hbox{\tiny id}}[\rho]=\int_Vd{\bm r}\rho({\bm r})
\left\{
\log{
\left[\rho({\bm r})\Lambda^3\right]
}-1
\right\},\nonumber\\\nonumber\\
&& \beta{\cal F}_{\hbox{\tiny ex}}[\rho]=-\frac{1}{2}\int_Vd{\bm r}_1\rho({\bm r}_1)\int_Vd{\bm r}_2\rho({\bm r}_2)f_{\hbox{\tiny M}}
({\bm r}_{12}),
\label{f3-4}
\end{eqnarray}
where $\Lambda$ is the thermal wavelength,
$f_{\hbox{\tiny M}}({\bm r}_{12})=\exp{\left[\beta u({\bm r}_{12})\right]}-1$ is the Mayer-function, 
$u({\bm r}_{12})$ is the pair potential between two particles, and ${\bm r}_{12}={\bm r}_1-{\bm r}_2$. In the case of hard cylinders 
the Mayer-function can be written as $f_{\hbox{\tiny M}}({\bm r}_{12})=-\Theta(D-R)\Theta(L-z)$, where $\Theta(x)$ is the Heaviside 
step function. Note that the ideal contribution is exact, while the excess part is approximated at the level of Onsager second-virial 
theory \cite{24}. To obtain the equilibrium density profile, the grand-potential functional has to be minimized with respect 
to the local density, i.e. $\delta\Omega/\delta\rho({\bm r})=0$. Using Eqns. (\ref{f1}) and (\ref{f3-4}), one can easily 
show that the equilibrium profile $\rho({\bm r})$ is given by the Euler-Lagrange equation
\begin{eqnarray}
\rho({\bm r})=e^{\displaystyle\beta\left[\mu-V_{\hbox{\tiny ext}}({\bm r})\right]}
\times e^{\displaystyle\int_Vd{\bm r}_1\rho({\bm r}_1)f_{\hbox{\tiny M}}({\bm r}-{\bm r}_{1})}.
\label{5}
\end{eqnarray}
As we focus our study on the structures of nematic and smectic phases in cylindrical confinement, the local density does 
not depend on the azimuthal angle $\phi$, but only on $R$ and $z$. Using the Mayer-function of hard cylinders and the external 
potential (\ref{f2}), the resulting Euler-Lagrange equation for $\rho(R,z)$, in the region where it is different from zero, 
is given by
\begin{eqnarray}
&&\rho(R,z)=e^{\displaystyle\beta\mu}
\exp\left[-\int_{z-L}^{z+L} dz_1\int_{\hbox{\tiny max}(0,R-D)}^{\hbox{\tiny min}
(R_e,R+D)} dR_1 R_1\right.\nonumber\\
&&\left.\times\rho(R_1,z_1)\Phi_{\hbox{\tiny exc}}\left(R,R_1\right)\right],
\label{f6}
\end{eqnarray}
where
\begin{eqnarray}
\Phi_{\hbox{\tiny exc}}\left(R_1,R_2\right)=\left\{\begin{array}{ll}
\displaystyle 2\pi,& R_{12}^+<D,\\
\displaystyle 2\cos^{-1}(t_{12}),&
R_{12}^-\leq D\leq R_{12}^+\\
\displaystyle 0,&R_{12}^->D,\end{array}\right.
\end{eqnarray}
where we have defined $t_{12}=\left(R_1^2+R_2^2-D^2\right)/\left(2R_1R_2\right)$, $R_{12}^-=|R_1-R_2|$ and $R_{12}^+=R_1+R_2$.
We have solved Eqn. (\ref{f6}) through the discretization of the local density, assuming periodicity along $z$ axis, using 
a trapezoidal rule for the numerical integration and an iterative method. The discretization is 
performed as follows: $\rho_{ij}\equiv\rho(R_i,z_j)$, where $i$ and $j$ are integers that define the discrete set of points 
in the pore, $R_i=i\Delta R$, with $\Delta R=R_e/N_R$, $i=0,\cdots,N_R$, and $z_j=j\Delta z$, with $\Delta z=d/N_z$, $j=0,\cdots,N_z$.
$d$ is the period of the smectic phase along the $z$ axis. In our calculations we used $\Delta z=0.05 L$ and $\Delta R=0.01 D$.
The discrete version of the equation is solved for given values of chemical potential $\mu$ and smectic period $d$, 
and the average packing fraction and smectic order parameter are determined from
\begin{eqnarray}
&&\eta_{\hbox{\tiny av}}=\frac{2\pi}{dA_e}\int_0^d dz\int_0^{R_e}dR R \eta(R,z),\nonumber\\\nonumber\\
&&S=\frac{2\pi}{dA_e}\int_0^d dz\int_0^{R_e}dR R \eta(R,z)\cos\left({\frac{2\pi z}{d}}\right),
\label{int}
\end{eqnarray}
where $\eta(R,z)=(\pi LD^2/4)\rho(R,z)$ is the local packing fraction, and $A_e=\pi R_e^2$ is the effective pore area. 
The integrals over $z$ and $R$ in Eqn. (\ref{int}) (and in fact all integrals in the present calculation)
are also computed using a trapezoidal rule. In the nematic phase the local density depends only on $R$, and 
\begin{eqnarray}
&&\rho(R)=e^{\displaystyle\beta\mu}\exp\left[-2L\int_{\hbox{\tiny max}(0,R-D)}^{\hbox{\tiny min}
(R_e,R+D)} dR_1 R_1\right.\nonumber\\
&&\left.\times \rho(R_1)\Phi_{\hbox{\tiny exc}}\left(R,R_1\right)\right],
\label{f9}
\end{eqnarray}
The smectic order parameter is zero for any solution $\rho(R)$ without $z$ dependence, while it has a positive value for a
smectic-like solution $\rho(R,z)$. We use the particle length $L$ as a scaling parameter for 
the smectic period, $d^*=d/L$, and the $z$ coordinate, $z^*=z/L$. The other scaling parameter is the particle diameter, $D$,
which we use to make $R$ and $W$ dimensionless: $R^*=R/D$ and $W^*=W/D$.
In the next section we present the nematic and smectic density profiles obtained from 
Eqns. (\ref{f6}) and (\ref{f9}) and determine the phase boundary of the nematic-smectic phase transition.

\section{Results and discussion}
\label{results}

\subsection{Structure of the fluid inside the pore}
\label{results_A}

In very narrow pores (1$<W^*<2$), there is no room for particles to overtake each other, and only nearest-neighbour interaction 
along $z$ occurs. The phase behaviour of the confined hard cylinders becomes identical to that of 
the one-dimensional Tonks gas \cite{27}. It is well-known that such a system does not exhibit a thermodynamic phase transition 
from fluid to crystal structures (note however that a very dense Tonks gas shows some changes from a fluid- to a solid-like 
structure, in the sense that the distance between the peaks of the positional distribution function is close to the average distance 
between neighboring particles; see the review of Giaquinta \cite{28}). The present approximate theory does predict a 
continuous fluid-crystal transition (the one-dimensional `crystal' corresponding to the smectic) at unrealistically high 
densities. Therefore one has to be cautious with the predictions of 
Onsager second-virial theory for high densities in very narrow pores. 

In the nematic fluid the density profile is homogeneous for $1<W^*<2$ because in this case $\Phi(R_1,R_2)=2\pi$ is a 
constant. The density profile derived from Eqn. (\ref{f9}) is
\begin{eqnarray}
\rho=e^{\displaystyle -2LA_e\rho+\beta\mu},
\label{f10}
\end{eqnarray}
which can be solved numerically for a given value of $\beta\mu$. This result is a consequence of the fact that the excluded 
volume of the particles, $V_{\hbox{\tiny exc}}=2LA_e$ is always the same regardless of the location of the rod (either in
the middle or at the wall of the pore). 

\begin{figure}
\epsfig{file=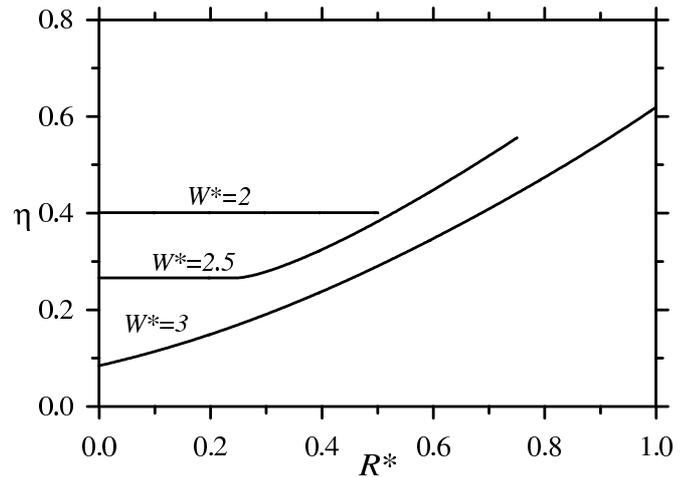,width=3.5in}
\caption{
 Dimensional crossover from the one-dimensional Tonks gas to the three-dimensional fluid. Density profiles 
of hard cylinders in very narrow pores at $\eta_{\hbox{\tiny av}}=0.4$.}
\label{fig2}
\end{figure}

This situation changes for $W^*>2$, because more than one cylinder may fit into the pore
in the same layer, i.e. overtaking can take place. The density profiles coming from the numerical solution of Eqn. (\ref{f9}) 
for $2\le W^*\le 3$ are presented in Fig. \ref{fig2}. One can see that the density profile changes continuously from a homogeneous to an
inhomogeneous distribution with increasing pore width. The reason why the density is constant in a small
interval of radial distances from the pore centre for the case $W^*=2.5$ is that particles staying in this region have the highest
excluded volume and do not allow any particle to overtake them. At the same time, the density peak at the wall is due to the fact 
that particles can reach minimal excluded volume at the wall, which makes it possible to maximize the free volume available for the 
rest of the particles. This manifests very clearly at $W^*=3$, where the fluid becomes completely inhomogeneous. 

\begin{figure}
\epsfig{file=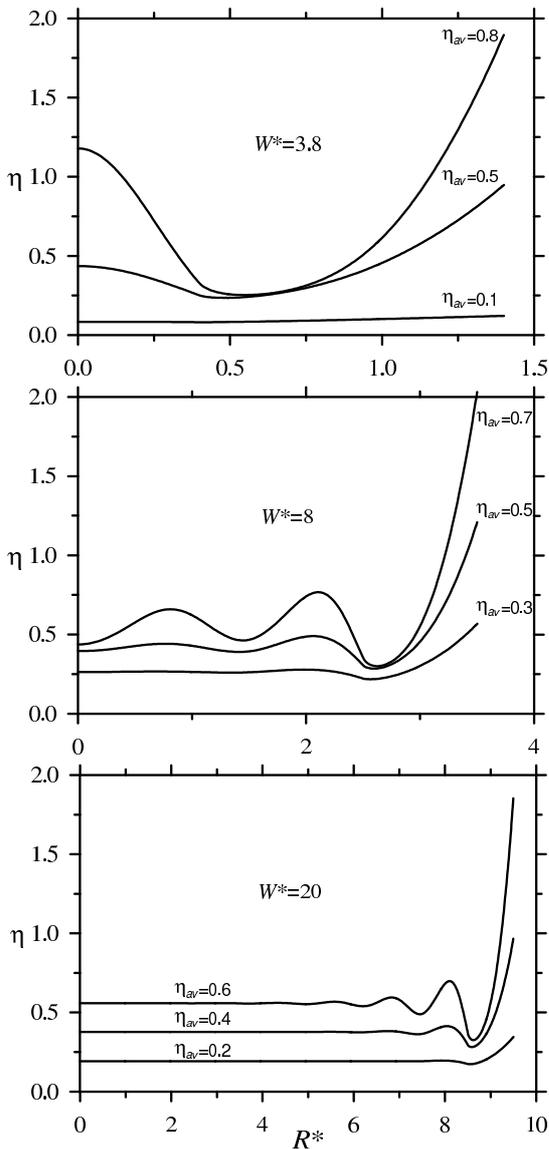,width=3.in}
\caption{
 Density profiles of hard cylinders as a function of average packing fraction $\eta_{\hbox{\tiny av}}$ 
in three different pores.}
\label{fig3}
\end{figure}

The structure of 
the fluid phase becomes even more interesting for wider pores, as shown in Fig. \ref{fig3}. The decreasing effect in excluded volume of the 
wall-particle interaction gives rise to oscillatory behaviour in the density profile. For $W^*=3.8$ the strong adsorption tendency 
of the particles at the wall results in a depletion zone between the wall and the pore central region, while the density 
peak at the centre and that at the wall are comparable. The propagation of the oscillatory behaviour to a wider range of distances can 
be observed with increasing density at $W^*=8$. For this pore width, the competition between wall-particle and particle-particle 
excluded-volume interactions results in a density minimum at the centre of the pore. The case $W^*=20$ demonstrates that the effect 
of the wall decays exponentially from the wall to the pore centre, with a typical decaying length not 
exceeding $\sim 3-4 D$ from the wall even at very high average packing fractions. Therefore the structure of a hard-cylinder 
fluid in very wide pores is altered only in the vicinity of the wall.

\subsection{Nematic-smectic transition in the pore}
\label{results_B}

The bulk system of parallel hard cylinders undergoes a continuous nematic-smectic phase transition. The nematic is a 
homogeneous fluid phase, while the smectic phase has a layered structure in the direction of the long axis of the cylinder (one-dimensional 
solid). The equations for the transition density  $\rho_{\hbox{\tiny NS}}^{\hbox{\tiny (bulk)}}$ and smectic period 
$d_{\hbox{\tiny NS}}^{\hbox{\tiny (bulk)}}$ of the bulk nematic-smectic transition are given by \cite{23}
\begin{eqnarray}
1+2\pi D^2\rho_{\hbox{\tiny NS}}^{\hbox{\tiny (bulk)}}\frac{
\sin\left({q_{\hbox{\tiny NS}}^{\hbox{\tiny (bulk)}}L}\right)}
{q_{\hbox{\tiny NS}}^{\hbox{\tiny (bulk)}}}
&=&0,\\
\frac{\partial}{\partial q_{\hbox{\tiny NS}}^{\hbox{\tiny (bulk)}}}
\left[\frac{
\sin\left({q_{\hbox{\tiny NS}}^{\hbox{\tiny (bulk)}}L}\right)}
{q_{\hbox{\tiny NS}}^{\hbox{\tiny (bulk)}}}
\right]&=&0
\label{f11}
\end{eqnarray}
where
$q_{\hbox{\tiny NS}}^{\hbox{\tiny (bulk)}}=2\pi/d_{\hbox{\tiny NS}}^{\hbox{\tiny (bulk)}}$ is the smectic wave number at the 
transition. These equations are obeyed by the values $\eta_{\hbox{\tiny NS}}^{\hbox{\tiny (bulk)}}\simeq 0.575$ and
$d_{\hbox{\tiny NS}}^{\hbox{\tiny (bulk)}*}\simeq 1.398$, where $\eta_{\hbox{\tiny NS}}^{\hbox{\tiny (bulk)}}=(\pi LD^2/4)
\rho_{\hbox{\tiny NS}}^{\hbox{\tiny (bulk)}}$ is the critical packing fraction. 
To determine the locus of the nematic-smectic transition points in the confined system, we add a smectic modulation to the 
fluid density as 
\begin{eqnarray}
\rho(R,z)=\rho_0(R)\left[1+\epsilon(R)\cos({qz})\right],
\label{f12}
\end{eqnarray}
where $\rho_0(R)$ is the solution of Eqn. (\ref{f9}) for a given $\beta\mu$ and $\epsilon(R)$ measures the amplitude of a small 
density modulation. After substitution of Eqn. (\ref{f12}) 
into Eqn. (\ref{f6}) and linearization in $\epsilon(R)$ of the exponential function, we obtain the following equation for 
$\epsilon(R)$:
\begin{eqnarray}
&&\epsilon(R)+2\frac{\sin{(qL)}}{q}\int_{\hbox{\tiny max}(0,R-D)}^{\hbox{\tiny min}(R_e,R+D)} dR' R'\rho_0(R')\nonumber\\
&&\times \Phi_{\hbox{\tiny exc}}(R,R') \epsilon(R')=0.
\label{f13}
\end{eqnarray}
Note that Eqn. (\ref{f13}) reduces to the first of Eqns. (\ref{f11}) in the limit $W\to\infty$, where the local density and 
$\epsilon(R)$ become constant. The case of very narrow pores ($1<W^*<2$) is very similar to the bulk case, because it is possible 
to simplify Eqn. (\ref{f13}) to
\begin{eqnarray}
1+2A_e\rho_{\hbox{\tiny NS}}\frac{\sin{(q_{\hbox{\tiny NS}}L)}}{q_{\hbox{\tiny NS}}}=0,
\label{f14}
\end{eqnarray}
where $\rho_{\hbox{\tiny NS}}$ satisfies Eqn. (\ref{f10}). Comparison of the first of Eqns. (\ref{f11}) and (\ref{f14})
shows that $\eta_{\hbox{\tiny NS}}=(\pi D^2/A_e)\eta_{\hbox{\tiny NS}}^{\hbox{\tiny (bulk)}}$.
To numerically determine $\rho_{\hbox{\tiny NS}}$ and $q_{\hbox{\tiny NS}}$ for other cases where $W$ is larger, it is useful 
to rewrite Eqn. (\ref{f13}) in discretised form, using the same space grid in radial direction $R$ introduced before:
\begin{eqnarray}
\sum_{k}{A}_{ik}\epsilon_k=0,
\label{f14a}
\end{eqnarray}
where ${A}_{ik}$ are the elements of a matrix $A$:
\begin{eqnarray}
{A}_{ik}=\delta_{ik}+2\frac{\sin{(qL)}}{q}c_k R_k\rho_0(R_k)\Phi_{\hbox{\tiny exc}}(R_i,R_k).
\end{eqnarray}
Here $\{c_k,R_k\}$ are weights and roots for the trapezoidal-rule integration [note that the $\{c_k\}$ roots take account
of the restricted integration interval in Eqn. (\ref{f13}), and give an $A$ matrix which is band-diagonal].
At the nematic-smectic transition, the ${A}$ matrix must obey the following two equations:
\begin{eqnarray}
\left.
\hbox{det}\hspace{0.1cm}\left({A}\right)
\right|_{\hbox{\tiny NS}}=0,\hspace{0.6cm}
\left.
\frac{\partial
\hbox{det}\hspace{0.1cm}\left({A}\right)}{\partial q}\right|_{\hbox{\tiny NS}}=0,
\label{f16}
\end{eqnarray}
which provide the transition density $\eta_{\hbox{\tiny NS}}$ and period $d_{\hbox{\tiny NS}}$. 
The dimension of the ${A}$ matrix depends on the values of
$R_e$ and $\Delta R$. For all the cases studied, with the chosen values of $R_e$ and $\Delta R$, the number of points used
result in accurate values for density and period.

One interesting outcome of Eqns. (\ref{f16}) is that the smectic period $d_{\hbox{\tiny NS}}$ does not depend on the pore-width and
is always equal to the bulk value, $d_{\hbox{\tiny NS}}=d_{\hbox{\tiny NS}}^{\hbox{\tiny (bulk)}}=1.398L$. This is 
due to the fact that the $R$ and $z$ variables are 
decoupled in the Mayer function. However, as far as the transition density is concerned, there must be a dependence with
pore width: in the bulk limit ($W\to\infty$), $\eta_{\hbox{\tiny NS}}$ should go to the bulk value 
$\eta_{\hbox{\tiny NS}}^{\hbox{\tiny (bulk)}}$, while it should diverge when $W\to D^+$ (note that, as mentioned in the beginning 
of Section \ref{results_A} in connection with the isomorphism between the present model and Tonks gas, 
there should be no transition for $W\le 2D$ and, as a consequence, 
the transition density should in fact diverge as $W\to 2D^+$). Therefore, the smectic phase should be destabilized with respect to the nematic phase with 
decreasing pore width.

\begin{figure}
\epsfig{file=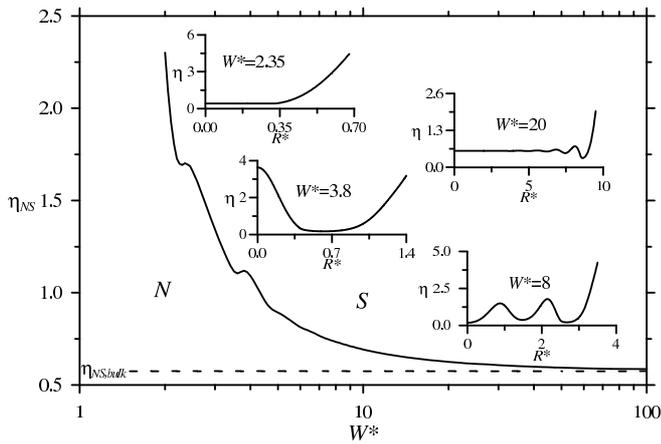,width=3.5in}
\caption{
Packing fraction $\eta_{\hbox{\tiny NS}}$ of the nematic-smectic phase transition as a function of pore width ($W$). 
Insets show a few density profiles along the nematic-smectic transition curve. The dashed line indicates the packing 
fraction $\eta_{\hbox{\tiny NS}}^{\hbox{\tiny (bulk)}}$ of the bulk nematic-smectic phase transition. $N$ and $S$ 
denote the regions of nematic and smectic phase stability, respectively.}
\label{fig4}
\end{figure}

The results of Eqn. (\ref{f16}) confirm this prediction. Fig. \ref{fig4} shows that the destabilization of the 
smectic phase is weak for very wide pores, while in narrow pores it gets very strong as the contribution of the 
wall-particle excluded volume becomes comparable to that due to particle-particle excluded volume. This effect, although
qualitatively correct, might be grossly overestimated by the present theory. For example,
for $W^*=5$, the increase in transition density exceeds the bulk value by 50\% (note that the resulting density is quite close 
to the close-packing limit of hard cylinders while it is known from MC simulations of hard cylinders \cite{19} that there is a 
large gap between the bulk nematic-smectic transition and the close-packing densities). Therefore it is clear that the predictions 
of our theory are correct but not quantitatively reliable for relatively narrow pores.

It is interesting to note the presence of damped oscillations in the 
nematic-smectic transition curve (see Fig. \ref{fig4}). This behavior is related to the 
commensuration between pore width and typical transverse particle distance when the former is changed. 
The density profiles along the nematic-smectic boundary shows very strong 
adsorption at the wall and depletion zones are also present at all pore widths. The structure in the pore centre is strongly 
affected by the pore diameter in narrow pores, since the structure generated at the surfaces may interfere coherently or incoherently in
the central region. By contrast, in wide pores the density modulation does not reach the pore centre and 
rapidly decays to an almost constant value (even at the nematic-smectic phase boundary), the inhomogeneous density being
restricted to a region of thickness $3-4 D$ from the wall (see the inset of $W^*=20$ in Fig. \ref{fig4}). 

\begin{figure}
\epsfig{file=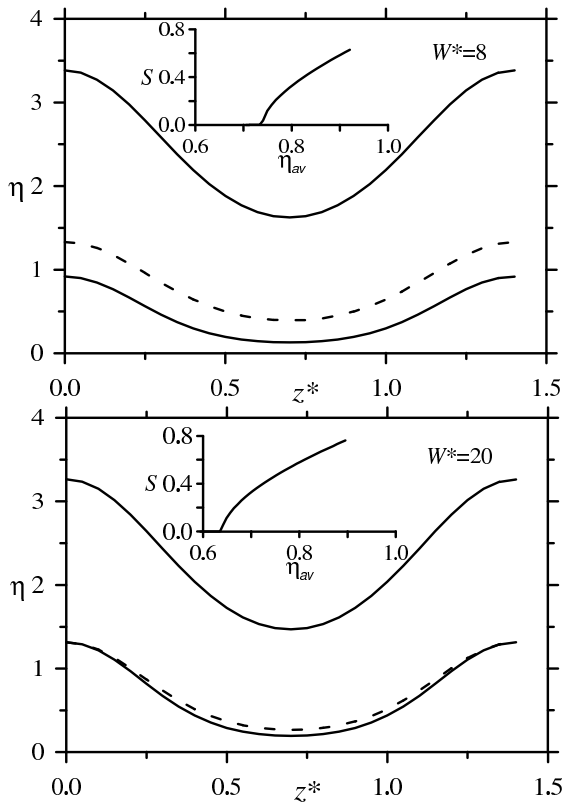,width=3.in}
\caption{
Density profiles of the smectic phase along the main axis of the pore at the wall (upper curves) and in the 
centre of the pore (lower curves). Top panel: $W^*=8$ and $\eta_{\hbox{\tiny av}}=0.8$. Bottom panel:
$W^*=20$ and $\eta_{\hbox{\tiny av}}=0.7$. The dashed curve is the average density profile along the main axis of the pore,
$\eta(z)=(2\pi/A_e)\int_0^{R_e} dR R \eta(R,z)$. The insets show the smectic order parameter $S$ as a function 
of average packing 
fraction $\eta_{\hbox{\tiny av}}$.} 
\label{fig5}
\end{figure}

Finally we show in Fig. \ref{fig5} the variation of the density profile of the smectic phase along 
the pore symmetry axis (with the value of the 
chemical potential chosen in the bulk smectic region). The upper curve shows the local density at the wall, $\eta(R_e,z)$, while 
the lower curve corresponds to the local density at the centre, $\eta(0,z)$. Note the smooth density modulations along the 
direction normal to the layers in both narrow and wide pores. 
In both cases the interstitial region ($z=d/2$) is not empty as the density is substantial even in the middle of
the pore, i.e. the one-dimensional ordering is not strong along the layer normal. The average density profile
along the pore axis (dashed curve) shows that the wall-particle excluded volume effect is very strong in a narrow pore, 
while it is marginal in a wide pore. This can be seen very clearly in the lower panel of Fig. \ref{fig5}, where the average density 
profile is just slightly different from the density profile in the pore centre. It can be also seen that the 
coupling in the ordering between normal and in-plane directions is weak in both cases. The insets of Fig. \ref{fig5} show
that the smectic order parameter goes smoothly to zero with decreasing packing fraction, i.e. the nematic-smectic phase 
transition is continuous even in cylindrical confinement. 

\section{Conclusions}

We have studied the effect of cylindrical confinement on the nematic and smectic phases of parallel hard rods using Onsager's 
second-virial theory. We found that the wall-rod interaction gives rise to inhomogeneous nematic and smectic structures, with 
strong adsorption taking place at the wall. The inhomogeneous packing of the rods gives rise to a decrease in
the average excluded volume of a particle, i.e. the free volume available for the particles in the pore can be increased 
substantially. In addition to this, the stability region of the smectic phase is shifted to higher densities with decreasing 
pore width. The observed nematic-smectic phase transition is continuous for any pore size. The reason for the 
destabilization of the smectic phase is that adding a one-dimensional spatial ordering to the inhomogeneous nematic phase 
cannot increase the packing entropy to the same extent as in the homogenous nematic phase. The transversal (in-plane) positional 
order is always present, while the longitudinal (out-of-plane) 
order takes place just above a critical density. In such a situation, the excluded volume gain from the 
longitudinal order cannot be as 
high as in the bulk nematic phase. It is worth mentioning that, in a fluid of Gay-Berne particles \cite{15},
the cylindrical confinement also hinders the formation of the smectic phase.

Although qualitatively correct, the predictions of this theory may be in error for narrow pores. A way to improve the predictions
(but certainly not the essential physics) is to use a Parsons-Lee scheme \cite{31,32}, where the high-density terms of the
diagrammatic expansion of the free-energy functional are approximated by those of a scaled fluid of hard spheres and the 
resulting series is resummed to give a density-renormalised free energy. This extended theory will, in the limit of small pore 
widths, give lower (and more realistic) values for the average packing fractions than those shown in Fig. \ref{fig4}. 

We have not examined the stability of possible columnar and solid-like structures of the confined hard cylinders. 
The stabilization of columnar ordering is quite unrealistic because it does not exist in the bulk limit and the commensuration 
conflict between pore-size and particle diameter does not support the formation of packed two-dimensional lattices in the pore.
However, the stabilization of columnar ordering cannot be excluded at some special pore widths, where the accommodation of the 
cylinders can be achieved efficiently in the pore. In very narrow pores, the existence of the nematic-smectic phase boundary is 
quite questionable and it may happen that a inhomogeneous nematic structure change continuously to a solid-like structure. The 
structure of such closely-packed phases may vary greatly with pore width, as is the case in cylindrically confined 
hard spheres \cite{29,30}. Our results clearly show that the nematic phase is the winner in the cylindrical confinement of
parallel particles. As regards the freely rotating case, we can say that the stability window of the nematic phase is expected 
to be even wider because confinement enhances the orientational ordering of hard rods \cite{13}. Therefore our study 
should be extended to the system of freely rotating hard rods in order to get a deeper understanding of the delicate interplay 
between the wall-rod and rod-rod interactions on the stability of isotropic, nematic, smectic and closed-packed structures.

\acknowledgments

SV acknowledges the financial support of the Hungarian State and the European Union under the TAMOP-4.2.2.A-11/1/KONV-2012-0071. 
Financial support from Comunidad Aut\'onoma de Madrid (Spain) under
the R$\&$D Programme of Activities MODELICO-CM/S2009ESP-1691, and from MINECO (Spain)
under grants FIS2010-22047-C01 and FIS2010-22047-C04 is also acknowledged.


\begin{thebibliography}{abcd}
\bibitem{1} S. Ishihara, J. Display Technol. {\bf 1}, 30 (2005).
\bibitem{2} P. Sheng, Phys. Rev. Lett. {\bf 37}, 1059 (1976).
\bibitem{3} R.-M. Marroum, G. S. Iannacchione, D. Finotello, and M. A. Lee, Phys. Rev. E {\bf 51}, 2743 (1995).
\bibitem{4} G. P. Sinha and F. M. Aliev, Phys. Rev. E {\bf 58}, 2001 (1998).
\bibitem{5} A. V. Kityk, M Wolff, K. Knorr, D. Morineau, R. Lefort and
P. Huber, Phys. Rev. Lett. {\bf 101}, 187801 (2008).
\bibitem{6} A. V. Kityk and P. Huber, Appl. Phys. Lett. {\bf 97}, 153124 (2010).
\bibitem{7} C. Grigoriadis, H. Duran, M. Steinhart, M. Kappl, H.-J. Butt and
G. Floudas, ACS Nano {\bf 5}, 9208 (2011).
\bibitem{8} C. V. Cerclier, M. Ndao, R. Busselez,R. Lefort,E. Grelet,P. Huber,
A. V. Kityk,L. Noirez, A. Sch\"onhals and D. Morineau, J. Phys. Chem. C {\bf 116}, 18990 (2012).
\bibitem{9} W.-Y. Zhang, Y. Jiang, and J. Z.Y. Chen, Phys. Rev. Lett. {\bf 108}, 057801 (2012).
\bibitem{10} S. Calus, D. Rau, P. Huber and A. V. Kityk, Phys. Rev. E {\bf 86}, 021701 (2012).
\bibitem{11} R. van Roij, M. Dijkstra and R. Evans, Europhys. Lett. {\bf 49}, 350 (2000).
\bibitem{12} M. Dijkstra, R. van Roij and R. Evans, Phys. Rev. E {\bf 63}, 051703 (2001).
\bibitem{13} D. de las Heras, E. Velasco and L. Mederos, Phys. Rev. Lett. {\bf 94}, 017801 (2005).
\bibitem{14} Q. Ji, R. Lefort and D. Morineau, Chem. Phys. Lett. {\bf 478}, 161 (2009).
\bibitem{15} Q. Ji, R. Lefort, R. Busselez and D. Morineau, J. Chem. Phys. {\bf 130}, 234501 (2009).
\bibitem{16} H. Reich and M. Schmidt, J. Phys.: Condens. Matter {\bf 19}, 326103 (2007).
\bibitem{17} M. M. Pineiro, A. Galindo and A. O. Parry, Soft Matter {\bf 3}, 768 (2007),
\bibitem{18} L. Harnau and S. Dietrich, Phys. Rev. E {\bf 66}, 051702 (2002).
\bibitem{19} J. A. C. Veerman and D. Frenkel, Phys. Rev. A {\bf 43} 4334 (1991).
\bibitem{20} J. A. Capit\'an, Y. Mart\'{\i}nez-Rat\'on and J. A. Cuesta, J. Chem. Phys. {\bf 128}, 194901 (2008).
\bibitem{21} L. Onsager, Ann. N. Y. Acad. Sci. {\bf 51}, 627 (1949).
\bibitem{22} G. J. Vroege and H. N. W. Lekkerkerker, Rep. Progr. Phys. {\bf 55}, 1241 (1992).
\bibitem{23} B. Mulder, Phys. Rev. A {\bf 35}, 3095 (1987); R. P. Sear and G. Jackson, J. Chem. Phys. {\bf 102}, 2622 (1995).
\bibitem{24} A. Malijevsky and S. Varga, J. Phys.: Condens. Matter, 22, 175002 (2010).
\bibitem{25} S. Varga, Y. Mart\'{\i}nez-Rat\'on and E. Velasco, Eur. Phys. J. E {\bf 32}, 89 (2010).
\bibitem{26} J. P Hansen and I. R. McDonald, Theory of Simple Liquids, 2nd ed. (Academic Press, London, 1986).
\bibitem{27} L. Tonks, Phys. Rev. {\bf 50}, 955 (1936).
\bibitem{28} P.V. Giaquinta, Entropy {\bf 10}, 248 (2008).
\bibitem{31} J. D. Parsons, J. Chem. Phys. {\bf 19}, 1225 (1979).
\bibitem{32} S. D. Lee, J. Chem. Phys. {\bf 87}, 4972 (1987).
\bibitem{29} M. C. Gordillo, B. Mart\'{\i}nez-Haya and J. M. Romero-Enrique, J. Chem. Phys. {\bf 125}, 144702 (2006); 
F. J. Dur\'an-Olivencia and M. C. Gordillo, Phys. Rev. E {\bf 79}, 061111 (2009)
\bibitem{30} A. Mughal, H. K. Chan and D. Weaire, Phys. Rev. Lett. {\bf 106}, 115704 (2011).
\end{thebibliography}
\end{document}